\documentstyle[twocolumn,aps,prb]{revtex}
\begin{document}

\twocolumn[\hsize\textwidth\columnwidth\hsize\csname @twocolumnfalse\endcsname

\title {Superconducting instability in the Holstein--Hubbard model:
A numerical renormalization group study }
\author{C.--H. Pao} 
\address{Department of Physics, National Chung Cheng University,\\ 
  Chia--Yi, 621 Taiwan, R.O.C.}
\author{H.--B. Sch\"uttler}
\address{Center for Simulational Physics, Department of Physics \& Astronomy,\\
 University of Georgia, Athens, GA  30602 }
\date{Auguest 25, 1997}

\maketitle

\widetext
\begin{abstract}
We have studied the $d$-wave pairing--instability in the 
two-dimensional Holstein-Hubbard model at the level of a full
fluctuation exchange approximation which treats
both Coulomb and electron-phonon (EP) interaction diagrammatically
on an equal footing. A generalized numerical renormalization 
group technique has been developed to solve the resulting 
self-consistent field equations.
The $d$-wave superconducting phase diagram shows an optimal
$T_c$ at electron concentration
$\langle n \rangle \sim 0.9$ for the purely electronic
Hubbard system. The EP interaction suppresses the $d$-wave $T_c$
which drops to zero when the phonon-mediated on-site attraction
$U_p$ becomes comparable to the on-site Coulomb repulsion $U$.
The isotope exponent $\alpha$ is negative in this model and small 
compared to the classical BCS value $\alpha_{BCS}={1 \over 2}$
or compared to typical observed values in non-optimally 
doped cuprate superconductors.
\end{abstract}
\pacs{PACS numbers: 74.20.-z, 74.20.Mn, 63.20.kr, 71.10.-w }
]
\narrowtext

In recent years, growing experimental evidence has 
suggested that YBa$_2$Cu$_3$O$_7$ and, possibly, other cuprates are
$d_{x^2-y^2}$--superconductors \cite{scalapino95}.
Anti--ferromagnetic (AF) spin fluctuation (SF) exchange
has been proposed as a possible candidate mechanism for $d$--wave
pairing.  These AF spin fluctuation models
are based on the notion that short-range, dynamical
AF spin correlations, caused by the strong local Coulomb 
repulsion in the cuprates, may lead to a 
spatially extended pairing 
attraction.\cite{scalapino95,bick89a,bick89,pao94a}
Starting from purely electronic models, 
such as the Hubbard Hamiltonian, 
coupling to lattice vibrational degrees of freedom is usually 
neglected in this picture.
However, except near certain ``optimal'' doping concentrations,
many cuprates, including YBa$_2$Cu$_3$O$_7$, exhibit a quite noticeable
doping dependent isotope effect \cite{isoexp}. This indicates that 
electron--phonon (EP) interactions could be important and should be 
included in the theory.

The goal of the present paper is to study the 
competition between phonons and AF spin fluctuation exchange
by means of a self-consistent diagrammatic approach which explicitly
includes phonon renormalizations to the AF spin fluctuations
at the level of the effective interaction vertices.
We formulate a full fluctuation exchange
(FLEX) \cite{bick89a,bick89} approximation which treats Coulomb and
EP contributions to the electron-electron interaction potential
entirely on an equal footing. Our work goes substantially beyond 
previous treatments \cite{mars90,schu95}
which have included phonon effects only at the 
level of the one--particle self--energy.

Due to the retarded nature of the EP interaction, the problem is 
numerically not directly amenable to the recent fast Fourier 
transform (FFT) \cite{sere91} or numerical 
renormalization group 
(NRG) \cite{pao94} methods, developed for 
for the FLEX approximation to the pure Hubbard model.
The present FLEX equations require certain large-scale 
fermion frequency matrix inversions which are numerically 
about 4 
orders of magnitude more demanding than FLEX calculations 
with short-range instantaneous interactions.
The numerical solution of this problem can
be achieved only by means of a generalized, highly efficient
matrix version of the original NRG method\cite{pao94}
which we have developed. 

We start from the simplest microscopic
Hamiltonian which includes both
an on--site Hubbard $U$ Coulomb repulsion and a local EP
coupling to an Einstein phonon branch,
the Holstein--Hubbard model \cite{schu92},
\begin{eqnarray}
  H &=& -t\sum_{\langle ij\rangle\sigma}\,
    \biggl [ c^{\dagger}_{i \sigma}
    c^{\vphantom{\dagger}}_{j \sigma} + HC \biggr ]\, -\,
    \mu\sum_{i\sigma}\, n_{i \sigma} \, +\, 
    U\sum_i\, n_{i \uparrow} n_{i\downarrow} \nonumber\\
    & & +\, \sum_{i}\, \left [{ {{\bf p}_{i}}^2 \over 2M}
        + {1 \over 2} K {{\bf u}_i}^2 \right ]\, -\,
        C \sum_{i\sigma}\thinspace {\bf u}_i
           \Big(n_{i \sigma} - {1 \over 2}\Big)\, ,
\end{eqnarray}
with a nearest neighbor hopping $t$, chemical potential $\mu$,
on--site Coulomb repulsion $U$, on--site EP coupling constant $C$, 
force constant $K$, and ionic oscillator mass $M$. 
The $c^{\dagger}_{i \sigma}$
($c^{\vphantom{\dagger}}_{i \sigma}$) is the electron creation (annihilation)
operator at site $i$ and spin $\sigma$; $n_{i \sigma}$ is the number operator;
and  ${\bf u}_i$ is the local ionic displacement at lattice site $i$.
The dispersionless bare phonon frequency is $\Omega_0\,=\, (K/M)^{1 / 2}$
and the phonon-mediated on--site attraction is $U_p\,=\,C^2 / K$.

Previous self-consistent field (SCF) studies \cite{mars90} of the 
Holstein-Hubbard system
have ignored the electron--electron exchange scattering which 
arises from the Pauli exclusion principle.
The importance of this exchange vertex\cite{bick89a,bick89,pao94a}
can be most easily demonstrated in the limit of the negative-U
Hubbard model with $U=-|U|$ and $U_p=0$.
In this case, the direct interaction
will give $2 U$ (after summing over the electron spin index)
and the exchange interaction
contributes $-U$. The simplest mean field theory will then predict the 
CDW instability to occur at 
$ | U | \bar{\chi}_{ph}(T)\, =\, 1$ 
(with exchange interaction) instead of 
$2 |U| \bar{\chi}_{ph}(T)\, =\, 1$ (without exchange interaction). 
In the positive--$U$ Hubbard model,
the exchange interaction enhances the spin fluctuations 
and thus helps the $d$--wave instability while at the same time
weakening the charge fluctuations.
Also, high phonon frequencies or a flat electron band
near the Fermi surface will tend to enhance the effect of the 
exchange vertex. To study particle--hole and 
particle-particle instabilities in this model,
it is thus necessary to include both Coulomb and phonon
contributions to the exchange vertex.

The bare interaction vertices for the FLEX equations shown in
Fig.\ref{vert}, include
the particle--hole
[Fig. \ref{vert}(b) and (c)] and the particle--particle [Fig. \ref{vert}(d)]
bare vertices, due to both the Hubbard $U$ and the 
phonon propagator 
$v_p(i \nu_m)\, =\, -{U_p \Omega^2_0  / (\Omega^2_0 + \nu_m^2)}$
[Fig. \ref{vert}(a)] for boson Matsubara frequency
$\nu_m = 2 m \pi T$.
The one--particle self--energy is then:
\cite {bick89,summation}
\begin{eqnarray}\label{eqnselfengy}
\Sigma_k&& =\ \sum_{k^\prime} 
    \Bigl [ V_2(\Delta k; i\omega_n )\, + \,  
   V^{ph}( \Delta k; i\omega_n) \Bigr ] G(k^\prime) \nonumber \\
   &&\ \ \ \ \ \ \ \ \ \ 
  + \ V^{pp}( \Delta k; i\omega_n) G^{*}(k^\prime) \ , 
\end{eqnarray}
\begin{eqnarray}\label{eqnv2}
V_2(\Delta k; i\omega_n )&&\, =\, -v_p(\Delta \omega)\, + \, 
  \sum_{i \omega_{n_1}}
  \Bigl [ v_p(\Delta \omega) + U \Bigr ] \nonumber \\
 &&\!\!\!\!\!\!\!\!\!\!\!\! \!\!\!\!\!\!\!\!\!\!\!\!\!\!\!\!\!\! 
 \Bigl [ 2 v_p(\Delta \omega)\! -\! 
   v_p(i\omega_n\! -\! i \omega_{n_1}\! -\! \Delta \omega)\! +\! U \Bigr ]
   \bar{\chi}_{ph} ( \Delta k; i\omega_{n_1})\ , 
\end{eqnarray}
\begin{eqnarray}
V^{ph}(\Delta k; i\omega_n )&&\ =\   \nonumber\\
 &&\!\!\!\!\!\!\!\!\!\!\!\!\!\!\!\!\!\!\!\!\!\!\!\!\!\!\!\!\!
\sum_{i \omega_{n_1}}
   {1 \over 2}
  \Bigl [ D(1+D)^{-1} - D \Bigr ]_{ n, n_1}\!\! (\Delta k)\ v^D_{n_1, n} 
     (\Delta \omega)\ + \nonumber \\ 
 &&\!\!\!\!\!\!\!\!\!\!\!\!\!\!\!\!\!\!
 { 3 \over 2} 
  \Bigl [ M(1+M)^{-1} - M \Bigr ]_{ n, n_1}\!\! (\Delta k)\ v^M_{n_1, n} 
     (\Delta \omega)\ , \label{eqnvph}\\ 
V^{pp}(\Delta k; i\omega_n )&&\  =\  \nonumber \\  
 &&\!\!\!\!\!\!\!\!\!\!\!\!\!\!\!\!\!\!\!\!\!\!\!\!\!\!\!\!\!
-\sum_{i \omega_{n_1}}
  \Bigl [ S(1+S)^{-1} - S \Bigr ]_{ n, n_1} \!\! (\Delta k)\ v^S_{n_1, n} 
     (\Delta \omega)\ + \nonumber \\ 
 &&\!\!\!\!\!\!\!\!\!\!\!\!\!\!\!\!
  3 \Bigl [ T(1+T)^{-1} - T \Bigr ]_{ n, n_1} \!\!(\Delta k)\ v^T_{n_1, n} 
     (\Delta \omega)\ ,   \label{eqnvpp} 
\end{eqnarray}
\begin{eqnarray}
 R_{n, n_1} (\Delta k) &&\ =\ v^R_{n, n_1}(\Delta \omega) \times \nonumber\\ 
 &&\!\!\!\!\!\!\!
   \Biggl \{ \begin{array}{l}
  \bar{\chi}_{ph}(\Delta k; i\omega_{n_1})\  \\ 
  \bar{\chi}_{pp}(\Delta k; i\omega_{n_1}) \end{array}
   \ {\rm for} \ R \, =\, \Biggl \{\begin{array}{l}
      D\ {\rm or}\ M\ ,\\    
      S\ {\rm or}\ T\ ,   
  \end{array} 
\end{eqnarray}
where 
$
k\,  \equiv\, ({\bf k},i\omega_n),\ \Delta k \, \equiv\, 
({\bf k - k^\prime}, i\omega_n - i\omega_{n^\prime}),
\ \omega_n=(2n+1)\pi T
$, 
the Green's function 
$
G(k)\, =\,
[ i\omega_n - \epsilon_{\bf k} - \Sigma_k ]^{-1}
$, 
and the tight binding band  
$
\epsilon_{\bf k}\, =\, -2t (\cos \, {\bf k}_x + \cos \, {\bf k}_y)
- \mu
$. 
The
$ v^R_{n, n_1}$ are the bare vertices shown in Fig. 1(b-d).
The bare particle--hole and particle--particle fluctuation functions 
are defined as:
\begin{figure}
   \includegraphics{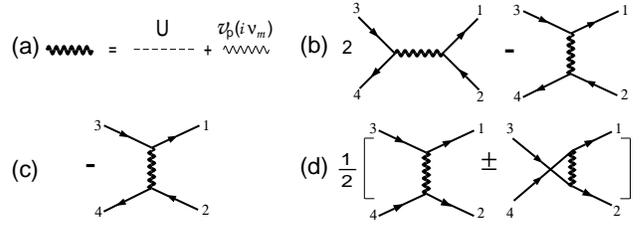}
\vspace*{1.5in}
\caption{The bare vertices of the Holstein--Hubbard model in the FLEX 
approximation. (a) The interactions of the on--site Coulomb $U$ and
Einstein phonon $v_p(i \nu_m)$. (b) Density vertex 
$v^D_{n_1,n_4}\!(i\nu_m)$ =
$[ 2 v_p(i \nu_m)\! -\! v_p(i\omega_{n_1}-i\omega_{n_4})\! +\! U]$ 
$\delta_{n_1,n_2+m} \delta_{n_3+m,n_4}$. (c) Magnetic vertex 
$v^M_{n_1,n_4}\!(i\nu_m)$ =
 $-[ v_p(i\omega_{n_1}-i\omega_{n_4})\! +\! U] $
$\delta_{n_1,n_2+m} \delta_{n_3+m,n_4}$. (d) Singlet and triplet vertices
 $v^{S}_{n_1,n_4}\!(i\nu_m)$ = $ 1/2 [ v_p(i\omega_{n_1}-i\omega_{n_4})$
 + $v_p(i\omega_{n_1}+i\omega_{n_4} -i\nu_m)\! +\! 2 U]$
 $\delta_{n_1,-n_2+m} \delta_{-n_3+m,n_4}$, 
 $v^{T}_{n_1,n_4}\!(i\nu_m)$ = $ 1/2 [ v_p(i\omega_{n_1}-i\omega_{n_4})
 \! -\! v_p(i\omega_{n_1}+i\omega_{n_4} -i\nu_m) ]$
 $\delta_{n_1,-n_2+m} \delta_{-n_3+m,n_4}$. }
 \label{vert}
\end{figure}
\begin{eqnarray}
\bar{\chi}_{ph}(q; i\omega_n) & = &  
   -{1 \over N} \sum_{\bf k} G(k+q) G(k) \ ,\\
\bar{\chi}_{pp}(q; i\omega_n) & = &\ \  {1 \over N} \sum_{\bf k} G(k+q) G(-k) \ .
\label{eqnchibar}
\end{eqnarray}
Because of the retarded nature 
of $v_p(i \nu_m)$, the
bare vertices in Fig.1(b-d) depend 
explicitly on the internal frequency 
transfer and $i\omega_n$ can not be summed out here.
The numerically most challenging part of the SCF
calculation is thus the evaluation of the fluctuation potentials,
$V^{ph}$ and $V^{pp}$, because of the required fermion
frequency matrix inversion in Eqs. (4,5).
In a brute--force approach, this matrix dimension
can become as large as $500 \times 500$ (the size
of the entire fermion Matsubara frequency set) 
at $T_c$. Recent FFT and NRG techniques for the pure Hubbard FLEX 
equations are not directly applicable and we had to
develop a generalized "fermion matrix" NRG method to handle
the numerics efficiently \cite{pao_up}.

Building upon the original NRG method \cite{pao94}, our
generalized fermion 
matrix NRG employs a frequency--RG operation which 
separates the frequency space into high and low regions 
at each temperature in such a way
that fermion frequency matrix dimensions will not increase
as fast as $T^{-1}$ when $T$ is lowered.
For a typical $8\times8$
matrix dimension, starting at some high 
temperature and large fermion frequency cut--off, 
we can keep the dimension
below 30 $\times$ 30 after 6 factor-2
frequency--RG steps, corresponding to a
temperature reduction from 4.0$t$ to 0.0625$t$. 
The fermion cut--off in the frequency RG steps
changes from 100$t$ ($\sim12\times$ bandwidth) to 1.6$t$ 
($\sim50\times T_c$). 
We then keep the 
same cut--off (without any further RG operation) and decrease the 
temperature slowly until the $d$--wave instability occurs \cite{pao94}. 
A typical matrix dimension is then 
about 50 $\times$ 50 near the $d$--wave instability. The $d$--wave 
instability is determined by solving the eigenvalue problem for the 
singlet particle--particle kernel, constructed from 
the full interaction potential in Fig. \ref{vert}(a),
\begin{figure}
 \includegraphics{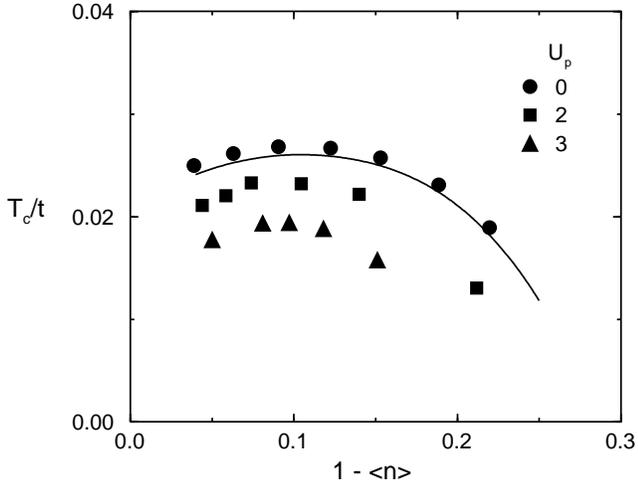}
\vspace*{2.7in}
\caption{Phase diagram of the 
$d$--wave instability with
$U/t = 4$ and $\Omega_0/t = 0.5$ and different $U_p$
on a 32$^2$ lattice.
Solid line represents a
Hubbard model calculation using the NRG method in 
Ref. [9]
which does $not$ involve the matrix inversion in Eqs. (4,5).  } 
\label{phase}
\end{figure}
\hspace*{-0.14in}following Ref. \onlinecite{bick89}. 
$T_c$ is reached when the maximum paring eigenvalue
$\kappa_d(T)=1$. A 32$\times$32 ${\bf{k}}$-mesh covering the full 
1st Brillouin zone has been employed in all of the calculations 
reported below.

Figure \ref{phase} shows the $d$--wave superconducting phase
diagram of the Holstein--Hubbard model with intermediate 
Hubbard $U/t = 4$ for various EP
coupling $U_p=0$, 2, and 3 and an Einstein phonon frequency 
$\Omega_0/t = 0.5$. 
Increasing the electron concentration $\langle n\rangle$
towards half--filling ($ \langle n \rangle \sim 1$) 
initially enhances the $d$--wave
$T_c$ until it reaches a maximum around 
$ \langle n \rangle \sim 0.9$. 
Beyond that point, at hole doping $x=1- \langle n \rangle $ 
below $\sim8\%$,
the strong AF fluctuations actually reduce the $d$--wave
$T_c$. This should be contrasted with early
Hubbard model FLEX results in Ref.
\onlinecite{bick89a}, where the detailed shape of the 
magnetic--superconducting boundary
was not well determined and the 
$d$--wave $T_c$ calculation was stopped when
the magnetic eigenvalue exceeded unity. 
Here, we have used a finer lattice mesh and larger
cut--off frequency and carefully pushed the 
$d$--wave calculations toward smaller hole doping.

In order to get a deeper understanding of the origin for
the $T_c$ maximum in Fig. 1,
we have carried out a McMillan-type
analysis\cite{mcmill} 
of the underlying pairing equations \cite{bick89} by estimating
the dimensionless Eliashberg parameter $\lambda_d$, which measures
the pairing potential strength averaged over the
Fermi surface in the $d$-wave channel,
and $\lambda_Z=-\partial_\omega{\rm Re}\Sigma(k)|_{\omega=i0^+}$, 
which measures the strength of the quasi-particle
mass enhancement, as well as the pair--breaking strength 
$\gamma=\Big|{\rm Im}\Sigma(k) \Big|_{\omega=i0^+} /T_c$,
due to the quasi-particle damping. 
Expressed in terms of "renormalized" parameters\cite{mcmill}
$\lambda_d^*=\lambda_d/(1+\lambda_Z)$
and $\gamma^*=\gamma/(1+\lambda_Z)$,
our results show that
\begin{figure}
  \includegraphics{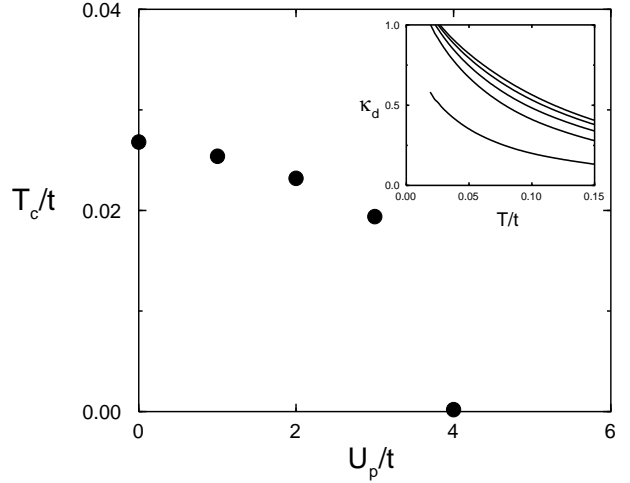}
\vspace*{2.7in}
\caption{
$d$--wave $T_c$ vs. EP potential $U_p$
for $U/t = 4$, $\langle n \rangle = 0.9$, and 
$\Omega_0/t = 0.5$. Inset is the maximum $d$--wave eigenvalue 
vs. $T$ for $U_p/t = 0, 1, 2, 3,\, {\rm and}\, 4$ 
(from top to bottom). 
$T_c$ for $U_p/t = 4$ is extrapolated
from the $\kappa_d$ data at the lowest available $T$. } 
\label{eigen}
\end{figure}
\hspace*{-0.14in}both the pairing strength $\lambda_d^*$
and the pair--breaking strength $\gamma^*$ are monotonically 
increasing as $\langle n \rangle$ is pushed towards 
half--filling. Thus there are (at least) two competing effects 
at work here: While raising $\lambda_d^*$ increases
$T_c$, raising $\gamma^*$ decreases it. Apparently,
for overdoping, the doping variation of the pairing strength
$\lambda_d^*$ dominates $T_c$, making $T_c$
initially rise with increasing $\langle n\rangle$.
On the other hand, for underdoping, close to half--filling, 
the doping variation of the pair--breaking strength $\gamma^*$
dominates and causes $T_c$ to decrease with increasing 
$\langle n\rangle$. An additional, related effect 
is that the overall AF spin fluctuation energy scale softens
as $\langle n\rangle$ approaches half-filling. 
This lowering of the relevant
"boson" energy scale will also lower $T_c$.

The primary effect of EP coupling is to suppress the 
$d$--wave $T_c$, shown in Fig. \ref{eigen} 
as a function of the EP potential strength $U_p$ (from 0 to 4t) 
at fixed $U/t = 4$, 
electron filling 
$ \langle n \rangle = 0.9$, 
and Einstein phonon frequency
$\Omega_0/t = 0.5$. 
The corresponding maximum $d$--wave pairing
eigenvalues $\kappa_d$ 
as a function of temperature are plotted in the inset.
Note that the $d$--wave $T_c$  drops to "almost zero"
(i.e. numerically inaccessible values) when 
$U_p$ becomes comparable to $U$.
This behavior is different from our earlier 
calculation\cite{schu95} 
which ignored the phonon renormalization of the bare 
interaction vertices.
In that case \cite{schu95}, $T_c$ was suppressed
only by the EP self-energy contribution, the suppression
was much more gradual and $T_c$ dropped only 
by about one half between $U_p = 0$ to $U_p \sim U$.
Here, by contrast, the EP interaction directly
counteracts the on-site Coulomb repulsion and thereby
suppresses the AF spin fluctuation mediated
pairing potential. The $d$--wave Eliashberg pairing
strength $\lambda_d$ and the pair--breaking strength $\gamma$
are indeed found to be strongly suppressed 
\nopagebreak by the EP interaction. 

\newpage
\vspace*{1.4in}
An important feature of EP coupling is that
it introduces an isotope effect into the 
electronic $d$--wave pairing mechanism.
Table I shows results for the isotope exponent
$\alpha = -d\log T_c/d\log M$ which becomes
$\alpha = {1 \over 2} d \log T_c /d \log \Omega_0$ in the 
present model, since the isotopic mass $M$
enters only through $\Omega_0$.
In our previous studies \cite{schu95} of the isotope effect,
where the EP effect on the magnetic bare vertices 
was neglected, the isotope exponent $\alpha$ was quite small and negative
for realistic phonon energies $\Omega_0$.
Here, as shown in Table I, we find qualitatively
the same result, even though the EP effect on the AF spin fluctuations
is now explicitly taken into account and suppresses $T_c$ 
much more strongly.
It is interesting to note that the absolute value of $\alpha$
has a minimum at the optimal doping concentration, a feature
qualitatively reminiscent of the doping dependent isotope data
in many cuprate systems.\cite{isoexp}
However the observed overall magnitude
of the effect in non-optimally doped cuprates, 
$|\alpha|\sim 0.5-1$, \cite{isoexp} is much larger than
the present model predicts.
This finding further supports the notion that the EP coupling
in the cuprates could be effectively very much enhanced
compared to conventional "strong-coupling" EP systems.\cite{schu95}

In conclusion, we have studied the $d$--wave superconducting instability
of the Holstein--Hubbard model in the FLEX approximation
by means of a generalized, matrix version of
the numerical renormalization group technique.
Upon including both the particle--particle
and particle--hole fluctuations, the $d$--wave $T_c$ shows a maximum at 
electron filling $\langle n \rangle \sim 0.9$. The $d$--wave $T_c$
is suppressed by increasing the EP potential $U_p$ and tends to zero
when $U_p \sim U$. Finally, the isotope 
exponents $\alpha$ are negative and small in magnitude, with typically
$|\alpha|<0.1$. While $|\alpha|$ exhibits a minimum at optimal
doping concentration, the overall magnitude is far too small to
explain observed isotope data in the cuprates. Our full FLEX 
results support the conclusions of earlier isotope 
calculations by the present authors. \cite{schu95}

\vspace{-7.7in}
\begin{table} 
\caption{Isotope exponent $\alpha$ for
$U/t = 4$ and $U_p/t = 2$.  }
\vspace{0.2in}
\label{table1}
\begin{tabular}{cdddd}
$\Omega_0$/t & 0.125 & 0.25 & 0.5 & 1.0 \\
\tableline
$\langle n \rangle$  \\
0.96 & -0.025 & -0.059 & -0.098 & -0.166 \\
0.90 & -0.022 & -0.053 & -0.090 & -0.127 \\
0.87 & -0.024 & -0.061 & -0.137 & -0.196 \\
\end{tabular}
\label{tabisotope}
\end{table}
\acknowledgments
We would like to thank Prof. N.E. Bickers for many
helpful discussions. This work was supported by the
National Science Council (Taiwan, R.O.C.) under Grant No.
862112-M194016T, by the National Science Foundation (U.S.A.) 
under Grant No. DMR-9215123 and by a
Nat'l Chung Cheng University Young Faculty Research Grant.
Computing support from UCNS (University of Georgia),
the Pittsburgh Supercomputer Center, and
the Nat'l Center for High--Performance Computing 
(NCHC, Taiwan, R.O.C.)
are acknowledged.


\end{document}